\documentclass[12pt]{article}
\usepackage{graphicx}
\usepackage{natbib} 
\usepackage{url} 

\usepackage{amsmath}
\usepackage{amssymb}
\usepackage{amsfonts}
\usepackage{multirow}
\usepackage{amsthm}

\usepackage{moreverb,url}

\usepackage{graphicx}
\usepackage{amsmath}
\usepackage{times}
\usepackage{bm}
\usepackage{enumerate}
\usepackage{mathtools}
\usepackage{setspace}
\usepackage{array,booktabs}
\usepackage{tikz}
\usepackage{natbib}
\usepackage{mathtools}
\usepackage{chngcntr}
\usepackage{algorithm}
\usepackage[utf8]{inputenc}
\usepackage{xcolor, soul}
\usepackage{hyperref}
\usepackage{multirow}
\usepackage{multicol}
\usepackage{subcaption}
\usepackage{amsmath}
\usepackage{footnote}

\DeclareMathOperator*{\argmin}{arg\,min}

\theoremstyle{definition}

\newcommand{\blind}{1}

\begin{document}

\def\spacingset#1{\renewcommand{\baselinestretch}%
{#1}\small\normalsize} \spacingset{1}

\date{}

\if1\blind
{
	\title{\bf Semiparametric Weighted Spline Regression (SWSR) in Confirmatory Clinical Trials with Time-Varying Placebo Effects}
	\author{
		Tianyu Zhan\thanks{Corresponding author email address: \texttt{tianyu.zhan.stats@gmail.com}.} \\
		\footnotesize Data and Statistical Sciences, AbbVie Inc., North Chicago, IL, USA\\
		\\
		Yihua Gu \\
		\footnotesize Data and Statistical Sciences, AbbVie Inc., North Chicago, IL, USA}
	\maketitle
} \fi

\if0\blind
{

} \fi

\bigskip
\begin{abstract}
In confirmatory Phase 3 clinical trials with recruitment over the years, the underlying placebo effect may follow an unknown temporal trend. Taking a clinical trial on Hidradenitis Suppurativa (HS) as an example, fluctuations or variabilities are common in HS-related endpoints, mainly due to the natural disease characteristics, variations of evaluation from different physicians, and standard of care evolvement. The adjustment of time-varying placebo effects receives some attention in adaptive clinical trials and platform trials, but is usually ignored in traditional non-adaptive designs. However, under the impact of such a time drift, some existing methods may not simultaneously control the type I error rate and achieve satisfactory power. In this article, we propose SWSR (Semiparametric Weighted Spline Regression) to estimate the treatment effect with B-splines to accommodate the time-varying placebo effects nonparametrically. Our method aims to achieve the following three objectives: a proper type I error rate control under varying settings, an overall high power to detect a potential treatment effect, and robustness to unknown time-varying placebo effects. Simulation studies and a case study provide supporting evidence. Those three key features make SWSR an appealing option to be pre-specified for practical confirmatory clinical trials. Supplemental materials, including the R code, additional simulation results and theoretical discussion, are available online. 
\end{abstract}

\noindent%
\vfill
\noindent%

\newpage
\spacingset{2} 

\section{Introduction}

A typical confirmatory Phase 3 clinical trial usually has a long study duration. For example, the median trial duration of over $20,000$ Phase 3 trials from 2000 to 2015 is $3.8$ years \citep{wong2019estimation}. During such a long period, time-varying placebo effects (or referred to as temporal trends, patient drift in some works) are very likely to occur, due to reasons such as evolving
standard of care, seasonal effects, resistance, and changes in viral species \citep{proschan2020resist}. In the BATTLE-1 trial on non-small cell
lung cancer (NSCLC) with recruitment from 2006 to 2009, more smokers were enrolled in the latter part as compared to the beginning of the study \citep{liu2015overview}. In the case study of Hidradenitis Suppurativa (HS), placebo data from several clinical trials spanning 2 years showed high fluctuations in responses. Such variabilities are common in HS-related endpoints, mainly due to the natural disease characteristics, variations of evaluation from different physicians, etc \citep{frew2020quantifying, garg2024addressing}. It can be challenging to justify that the underlying placebo effect is constant or follows a known parametric model. 

This topic has received some attention in platform trials \citep{roig2022model, wang2023bayesian}, and adaptive clinical trials, especially response-adaptive randomization designs \citep{korn2011outcome, villar2018response, proschan2020resist, robertson2023response}. However, even for non-adaptive standard Phase 3 clinical trials, we still need advanced analysis methods to handle time-varying placebo effects with at least the following three properties: (1) proper type I error rate control; (2) high power to detect a potential treatment effect; (3) robustness to unknown non-linear time-varying placebo effects. 

There are some existing methods available to handle this problem. For example, randomization tests \citep{rosenberger2015randomization, proschan2019re, simon2011using} that treat observed data as fixed and regenerate allocations under the null hypothesis is a robust method to control type I error rates, but may come with the cost of a reduced power \citep{villar2018response, robertson2023response}. Another approach is to incorporate time information in the regression analysis based on certain model assumptions \citep{robertson2023response}, e.g., linear functions \citep{coad1991sequential}, polynomial functions \citep{jennison2001group}, parametric forms in the logistic regression \citep{villar2018response}. However, type I error rates and power of those methods may not be robust to unknown time-varying placebo effects, which can deviate from the assumed parametric form in practice. Therefore, a new method with those three features is still necessitated. 

In this article, we propose a framework called SWSR (Semiparametric Weighted Spline Regression) to make statistical inferences of the treatment effect with a parametric form, and to model the time-varying placebo effects nonparametrically by B-splines \citep{de1978practical, hastie1990generalized, marsh2001spline, fahrmeir2013regression}. The workflow of SWSR is streamlined in Algorithm \ref{alg_SSR} with more details in Section \ref{sec:prop}. Our SWSR has several features to be highlighted: (1) It can properly control the type I error rate at the nominal level, and can accurately estimate the treatment effect with an accurate coverage probability, especially under heteroscedasticity with unequal variances; (2) SWSR has an overall better power performance and a smaller standard error of the treatment effect estimation; (3) SWSR is robust to model misspecification of the time-varying effects. Based on simulation studies in Section \ref{sec:sim}, SWSR has the overall best performance across all scenarios evaluated. It is appealing in practice to make SWSR a pre-specified analysis method at the study design stage or before the unblinding process.  

The remainder of this article is organized as follows. In Section \ref{sec:setup}, we provide a setup of the problem. Then we review some existing methods and introduce our proposed SWSR in Section \ref{sec:method}. Simulation studies are conducted to demonstrate the advantages of our SWSR versus competitors in Section \ref{sec:sim}. In Section \ref{sec:real}, we apply our method to a case study. Some discussions are provided in Section \ref{sec:dis} 

\section{Time-varying Placebo Effects}
\label{sec:setup}

Consider a two-group randomized clinical trial of a total sample size $N$ with placebo $p$ and treatment $t$. Let $a_i$ denote the treatment indicator for patient $i$ ($i = 1, ..., N$), taking a value of $1$ if randomized to treatment, and $0$ if randomized to placebo. We assume that the response $y_{i}$ for patient $i$ follows a Normal distribution, 
\begin{equation}
\label{equ:assump}
y_{i} \sim \mathcal{N}\left(\mu_{i}, \sigma_i^2\right),
\end{equation}
where $\sigma_i$ is equal to $\sigma_p$ if $a_i = 0$, and $\sigma_t$ if $a_i = 1$. The standard deviations $\sigma_i$'s within the same group are the same. The mean parameter $\mu_{i}$ in (\ref{equ:assump}) is defined as
\begin{equation}
\label{equ:mean}
\mu_{i} = f(t_{i}) + a_i\theta, 
\end{equation}
where $t_{i}$ is the data collection time of $y_{i}$. The function form of $f(t)$ is unknown. The parameter $\theta$ is interpreted as the treatment effect or treatment difference between the treatment group $t$ and the placebo group $p$. Our objectives are to test if $\theta$ is larger than zero under a specified significance level $\alpha$, and to provide a point estimation of $\theta$, based on observed data $x = \left\{y_i, t_i, a_i\right\}_{i=1}^N$.

Here are some remarks on (\ref{equ:assump}) and (\ref{equ:mean}). First of all, the $f(t)$ in (\ref{equ:mean}) captures a time-varying effect on $\mu_{i}$. When $f(t)$ is a constant with respect to $t$ during trial duration or when $t_{i}$ is the same for all patients, $f(t)$ in (\ref{equ:mean}) represents a constant placebo effect. This problem degenerates to a traditional one of testing means from two Normal distributions with unequal variances, in which, the Welch's $t$-test has a satisfactory performance \citep{lehmann2006testing}. However, in practical clinical trials with a moderate or long enrollment time, $f(t)$ is not always a constant due to variation in the standard of care and other reasons, and $t_{i}$'s are different between patients. Secondly, $f(t)$ has an unknown and complex non-linear functional form, and is also a nuisance parameter when performing hypothesis testing of $\theta$. The linear relationship between $\theta$ and $f(t)$ in (\ref{equ:mean}) is motivated by the interpretation of $\theta$ as the treatment difference. This parametric term can be generalized to other forms, e.g., odds ratio, in other problems. Thirdly, we focus on (\ref{equ:mean}) in this article. It can be generalized to include additional covariates in $f(t)$ and additional parametric components (e.g., linear terms) next to $\theta$. There is no assumed $f(t)$ by $a$ interaction in (\ref{equ:mean}). This article focuses on a basic setting that the temporal covariate $f(t)$ is a prognostic variable that is associated with outcome but not $a$ \citep{fda2023}. Its generalization is discussed in Section \ref{sec:dis}. 

\section{Methods}
\label{sec:method}

\subsection{Existing Methods}
\label{sec:comparator}

We first consider a starting point where $f(t)$ is known, and therefore $f(t)$ can be treated as a known offset in modeling. The homoscedasticity assumption of the simple linear regression (SLR) is not satisfied, because of the different $\sigma_p$ and $\sigma_t$ from two groups in (\ref{equ:assump}). In general, weighted linear regression (WLR) can be applied to handle heteroscedasticity in (\ref{equ:assump}). The weight for each observation is set as the inverse of variance $\sigma_i^2$ \citep{montgomery2021introduction}. Since $\sigma_i^2$ is unknown, one can first fit SLR on data, and then get $\widehat{\sigma}_i^2$ based on mean of squared residuals from the same group $p$ or $t$ \citep{shao2008mathematical}. Another approach is the robust regression (RR) using an M estimator \citep{huber2011robust} implemented by the R package $\texttt{MASS}$ \citep{ven2002}. These three methods (SLR, WLR and RR) can incorporate some parametric forms of $t_i$ in their models to account for time-varying placebo effects. However, power may not be preserved if this model assumption is misspecified, as shown in Section \ref{sec:sim}. 

When $f(t)$ is unknown with a non-linear functional form, the dimension of $f(t)$ will increase with the sample size $N$. Some direct methods, such as maximum likelihood estimation, are not applicable. Randomization tests are particularly useful where there is a time drift in the null hypothesis \citep{rosenberger2015randomization}, as the problem we considered in (\ref{equ:assump}) and (\ref{equ:mean}). Randomization tests usually treat observed response data $\left\{y_i\right\}_{i=1}^N$ as fixed, and then obtain the \textit{p}-value of a specific test statistic based on the empirical probability of observing more extreme data by shuffling the treatment assignments $\left\{a_i\right\}_{i=1}^N$ \citep{proschan2019re}. Those test statistics in randomization tests have a wide range of choices, such as mean difference, median difference, Welch's $t$-test statistic or Wilcoxon rank sum test statistic. \citet{simon2011using} had a discussion on conditions for randomization tests to have a proper type I error rate control. Simulation studies in Section \ref{sec:sim} show that a randomization test with a proper test statistic has a proper control of type I error rate, but our proposed method introduced in the next section has an overall better power performance. 

\subsection{A Proposed Method}
\label{sec:prop}

In this work, we propose SWSR (Semiparametric Weighted Spline Regression) to make statistical inference and hypothesis testing on $\theta$ in (\ref{equ:mean}) with a nonparametric form of $f(t)$ modeled by B-splines \citep{de1978practical}. The workflow of SWSR is streamlined in Algorithm \ref{alg_SSR}. 

We first provide a short review of the B-splines, with more details that can be found in \citet{de1978practical, hastie1990generalized, marsh2001spline, fahrmeir2013regression}. The following working model of $\widetilde{f}(t; k, d, \boldsymbol{\gamma})$ is constructed to estimate the non-linear function $f(t)$,
\begin{equation}
\label{eq:bs}
\widetilde{f}(t; k, d, \boldsymbol{\gamma}) = \sum_{j=1}^{k+d+1}\gamma_j B^d_j(t),
\end{equation}
where $k$ is the number of internal knots (excluding boundary knots), $d$ is the degree (e.g., $3$ for cubic B-splines), $\gamma_j$ is the $j$th coefficient of $\boldsymbol{\gamma}$ to be estimated, and $B^d_j(t)$ is the $j$th B-spline basis of order $d$ defined by recursive equations from degree $0$ up to $d$ with details in \citet{de1978practical}. 

\begin{figure}
	\begin{subfigure}{.5\textwidth}
		\centering
		\includegraphics[width=1\linewidth]{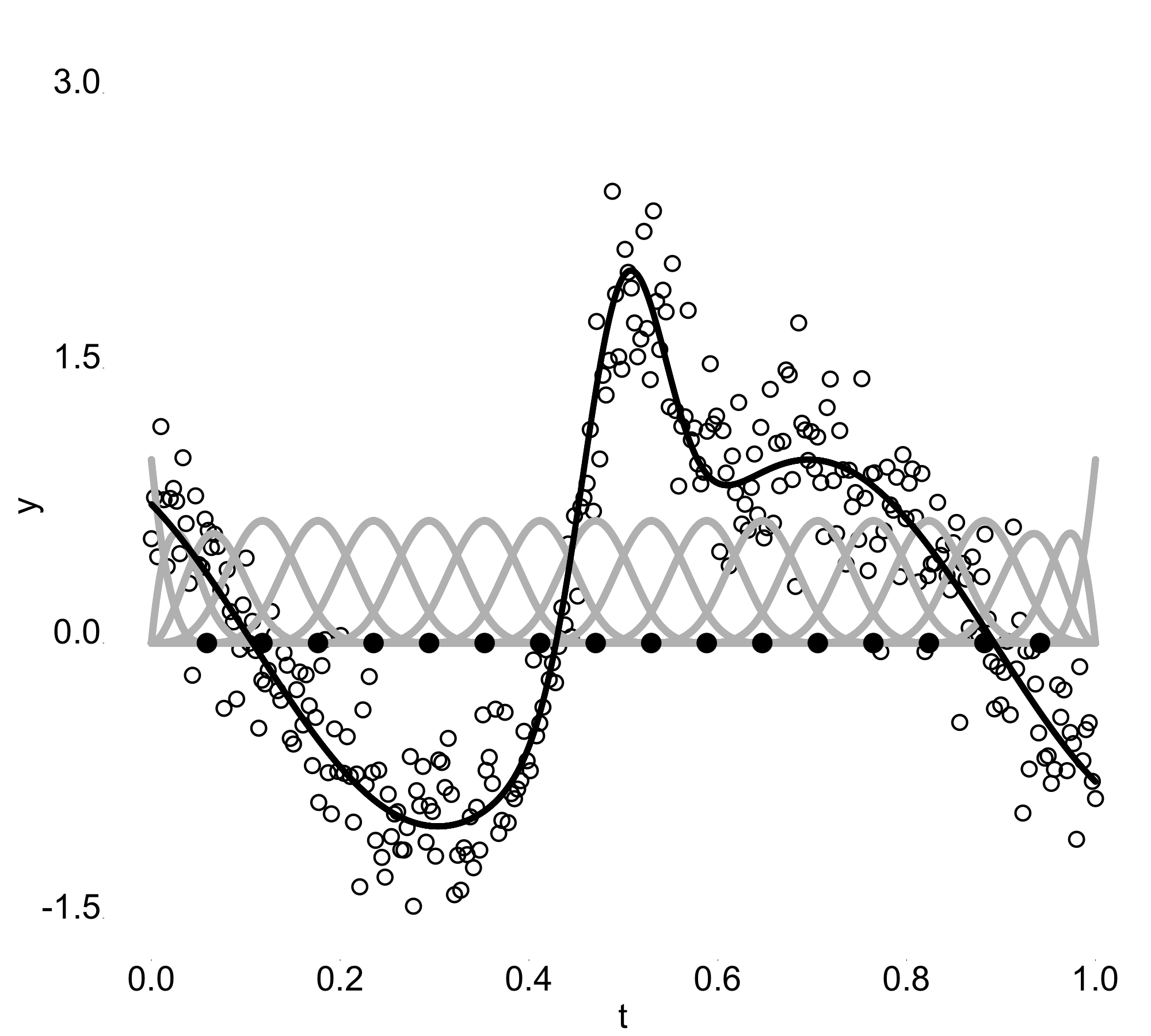}
		\caption{B-spline bases $\left\{B^d_j\right\}_{j=1}^{k+d+1}$}
		\label{fig11}
	\end{subfigure}%
	\begin{subfigure}{.5\textwidth}
		\centering
		\includegraphics[width=1\linewidth]{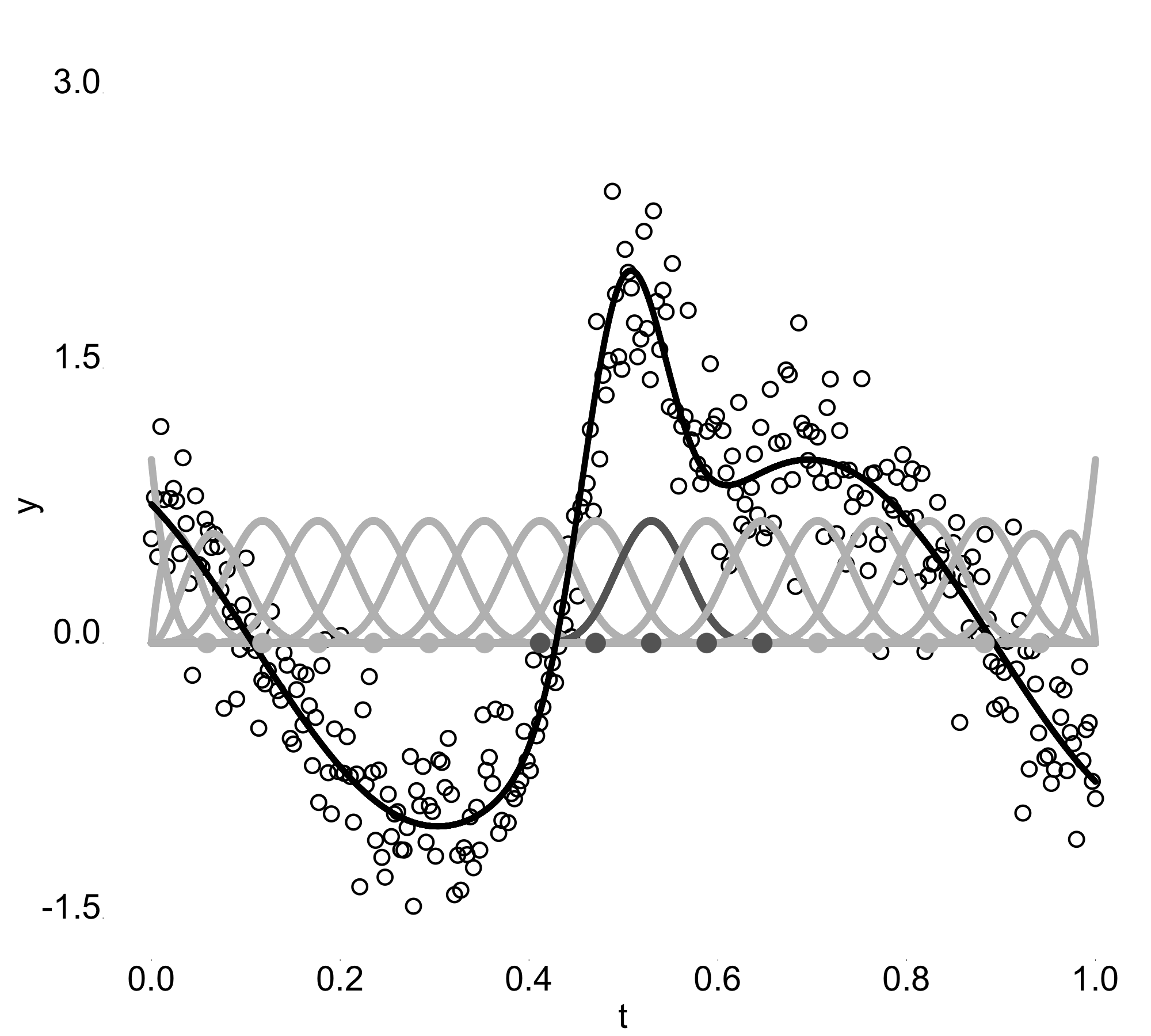}
		\caption{A B-spline basis spans at most $d+2$ internal knots}
		\label{fig12}
	\end{subfigure}
	\begin{subfigure}{.5\textwidth}
		\centering
		\includegraphics[width=1\linewidth]{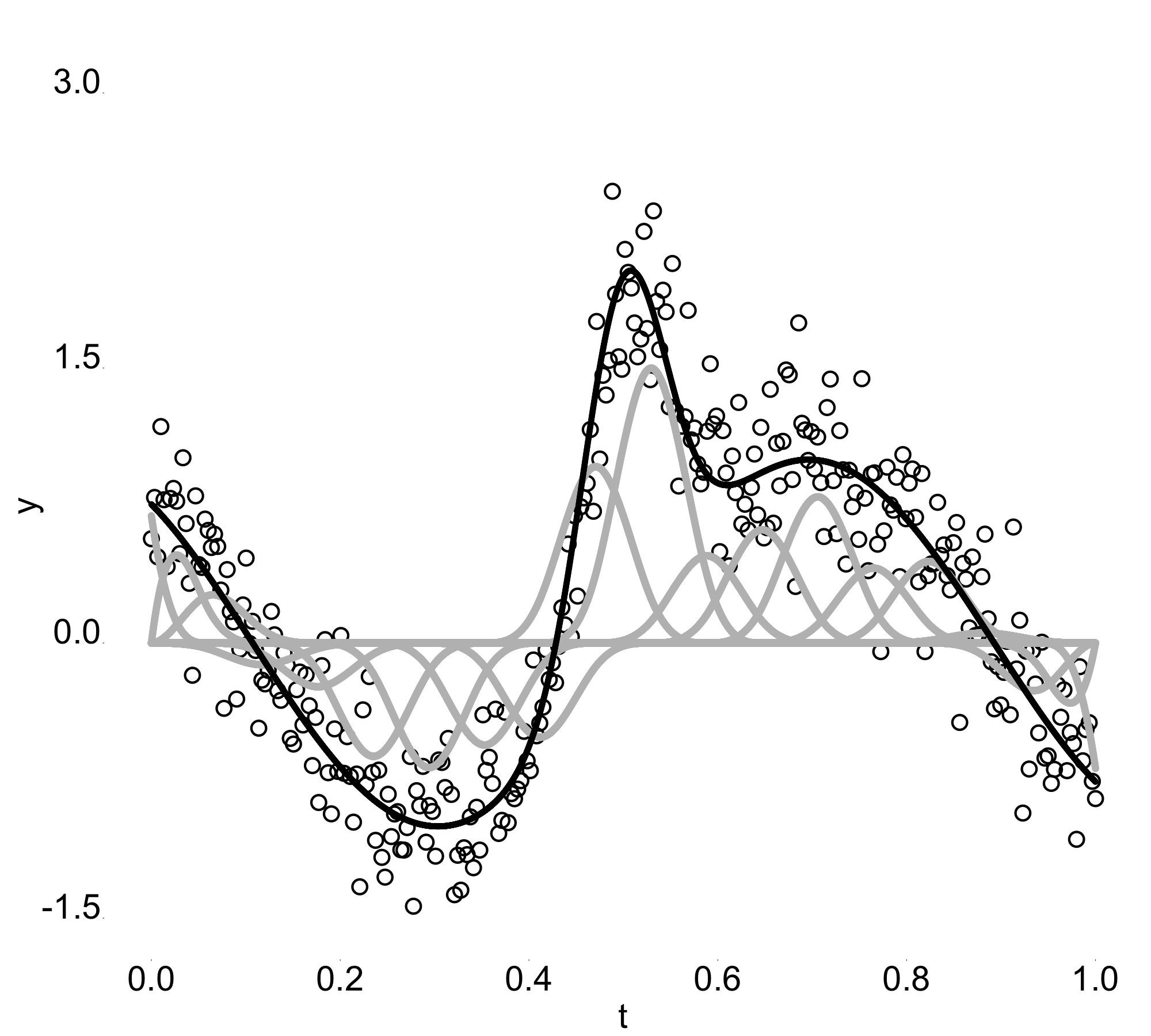}
		\caption{Scaled B-spline bases $\left\{\widehat{\gamma}_j B^d_j\right\}_{j=1}^{k+d+1}$}
		\label{fig2}
	\end{subfigure}%
	\begin{subfigure}{.5\textwidth}
		\centering
		\includegraphics[width=1\linewidth]{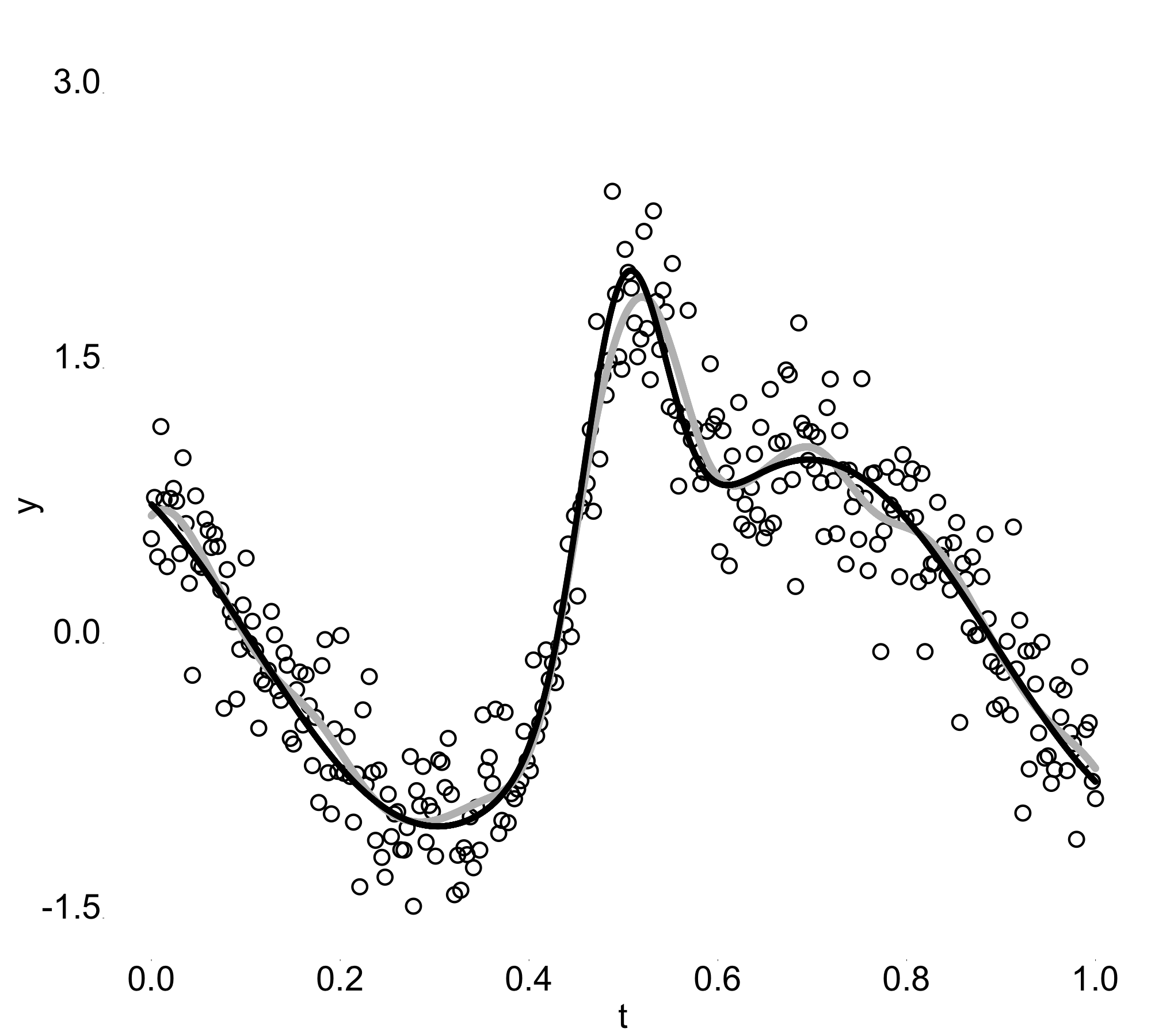}
		\caption{Approximate $f(t)$ in black by $\widetilde{f}(t; k, d, \widehat{\boldsymbol{\gamma}})$ in light gray}
		\label{fig3}
	\end{subfigure}
	\caption{Demonstration of B-splines $\widetilde{f}(t; k, d, \widehat{\boldsymbol{\gamma}})$ with $k=16$ internal knots and a degree of $d=3$ to approximate $f(t)$. Based on \citet{fahrmeir2013regression}, the $300$ data points are simulated from $y = f(t)+\epsilon$, where $f(t) = \sin\left[2\left(4t-2\right) \right] + 2 \exp \left[ -16^2\left(t-0.5 \right)^2 \right]$, and $\epsilon \sim \mathcal{N}(0, 0.09)$. }
	\label{fig_bs}
\end{figure}

In Figure \ref{fig11} as a demonstrating example based on \citet{fahrmeir2013regression}, the target unknown function $f(t)$ for $t \in [0, 1]$ is in a black line, with $300$ data points in hollow dots simulated from $f(t)$ with independent normal errors of a mean $0$ and a standard deviation $0.3$. We consider B-splines with $k=16$ internal knots labeled by black dots on the x-axis (excluding 2 boundary knots at $t=0$ and $t=1$) and a degree of $d = 3$. Therefore, there are $20 = 16+3+1$ B-spline bases shown in light gray lines in Figure \ref{fig11}. Each B-spline basis spans at most $d+2$ internal knots. For example, the dark gray B-spline basis in Figure \ref{fig12} covers $5 = 3 + 2$ internal knots.

Having B-spline bases $\left\{B^d_j\right\}_{j=1}^{k+d+1}$ set up, we can compute $\widehat{\boldsymbol{\gamma}}$ as the least squares estimate of $\boldsymbol{\gamma}$ in (\ref{eq:bs}). Figure \ref{fig2} shows scaled B-spline bases $\left\{\widehat{\gamma}_j B^d_j\right\}_{j=1}^{k+d+1}$ in light gray lines. The final $\widetilde{f}(t; k, d, \widehat{\boldsymbol{\gamma}})$ is calculated by summing up those scaled B-spline bases $\left\{\widehat{\gamma}_j B^d_j\right\}_{j=1}^{k+d+1}$, and it is close to the true unknown function $f(t)$ in a black line (Figure \ref{fig3}). B-splines enjoy those smoothing properties of continuous lower-degree derivatives at knots that are available in many other spline methods, e.g., the truncated power series \citep{hastie1990generalized, marsh2001spline}. Since a B-spline basis only spans at most $d+2$ internal knots, this method is also numerically more stable \citep{hastie1990generalized, fahrmeir2013regression}. However, one challenge of B-splines is to properly choose the values of $k$ and $d$. 

In Step 1 of Algorithm \ref{alg_SSR} for SWSR, 5-fold cross-validation is conducted to select a proper $k$ as the number of internal knots and a proper $d$ as the degree of B-splines. We consider a total of $C$ sets of hyperparameters in cross-validation with varying values of $k$ and $d$. In practice, $d \leq 3$ is usually sufficient for typical problems \citep{hastie1990generalized, fahrmeir2013regression}. The $k$ can range from $k=1$ for a simple model to a moderate value of $5$ or $10$, with internal knots as the $j/(k+1)$-quantiles (for $j = 1, ..., k$) of the observed time points $\left\{t_i\right\}_{i=1}^N$ \citep{fahrmeir2013regression}. More knots are placed in regions with a larger number of covariates. As can be seen from later simulation results in Section \ref{sec:sim}, this approach based on cross-validation is more stable than methods with a simple model or a relatively complex model. 

In Step 1.2 and 1.4.1 with given $k$ and $d$, we construct the following working model on $\left\{y_i\right\}_{i=1}^N$ based on (\ref{equ:mean}) and (\ref{eq:bs}),
\begin{equation}
\label{eq:work}
y_{i} = \sum_{j=1}^{k+d+1}\gamma_j B^d_j(t_i) + a_i\theta + \epsilon_i,
\end{equation}
where $\epsilon_i \sim \mathcal{N}(0, \sigma_i)$, $\sigma_i$ is equal to $\sigma_p$ if $a_i = 0$, and $\sigma_t$ if $a_i = 1$. Similar to WLR of handling unequal variances, we apply weights to the model fitting. Denote $w_i$ as the weight for the observation $i$, $i = 1, ..., N$. We compute the weighted least squares estimate $\widehat{\boldsymbol{\gamma}}$ and $\widehat{\theta}$ as,
\begin{equation}
\label{eq:wls}
\left(\widehat{\boldsymbol{\gamma}}, \widehat{\theta}\right) = \argmin_{\boldsymbol{\gamma}, {\theta}} \sum_{i=1}^N\left[w_i \left(y_i - \sum_{j=1}^{k+d+1}\gamma_j B^d_j(t_i) - a_i\theta \right) \right]^2.
\end{equation}
Since true variances are unknown, we first fit an unweighted model (\ref{eq:work}) on data in Step 1.2, and then calculate weights $\left\{{w}_i\right\}_{i=1}^N$ as the inverse of the mean of squared residuals in Step 1.3 from the same group $p$ or $t$. This process is similar to the WLR for model assumption in (\ref{equ:assump}), where observations within the same group have the same variance \citep{shao2008mathematical}. The weighted regression with model (\ref{eq:work}) is conducted with weights $\left\{{w}_i\right\}_{i=1}^N$ in Step 1.4.1. In Step 1.6, $k_s$ and $d_s$ for $s$ among $1, ..., C$ are selected with the smallest average validation MSE in cross-validation. 

In Step 2, we implement Steps 1.2, 1.3 and 1.4.1 with the selected parameters $k_s$ and $d_s$ on all data $x = \left\{y_i, t_i, a_i\right\}_{i=1}^N$ to compute $\widehat{\theta}$ in (\ref{eq:wls}) and its standard error (SE). 

\begin{algorithm}
	\caption{(SWSR) Semiparametric Weighted Spline Regression}
	\label{alg_SSR}
	\begin{tabbing}
		1. Conduct cross-validation to select $k$ and $d$ in $\widetilde{f}(t; k, d, \boldsymbol{\gamma})$. \\
		\quad 1.1. Generate 5-fold cross-validation samples. \\
		\quad 1.2. For given $k_c$ and $d_c$, $c = 1, ..., C$, fit an unweighted spline regression with the model (\ref{eq:work}). \\
		\quad 1.3. Compute the weight ${w}_i$ for patient $i$ as the inverse of the mean of squared residuals
		\\
		\quad\quad\quad from the same group $p$ or $t$. \\
		\quad 1.4. Within each training dataset, \\
		\qquad 1.4.1. Fit a weighted spline regression with the working model (\ref{eq:work}) and weights $\left\{{w}_i\right\}_{i=1}^N$. \\
		\qquad 1.4.2. Compute the mean squared error (MSE) of the weighted spline regression from the \\
		\qquad\qquad\: corresponding validation dataset.\\
		\quad 1.5. Calculate the average of MSEs from all validation datasets for given $k_c$ and $d_c$. \\
		\quad 1.6. Select $k_s$ and $d_s$ for $s$ among $1, ..., C$ with the smallest average validation MSE.\\
		\\
		2. Implement Steps 1.2, 1.3 and 1.4.1 with $\widetilde{f}(t; k_s, d_s)$ on all observed data $x = \left\{y_i, t_i, a_i\right\}_{i=1}^N$\\
		\:\:\: to obtain $\widehat{\theta}$ in (\ref{eq:wls}) and its corresponding standard error (SE). 
	\end{tabbing}
\end{algorithm}

\section{Simulation Studies}
\label{sec:sim}

In this section, we conduct simulation studies to evaluate the performance of the proposed SWSR and some comparators in a two-group clinical trial with a total sample size of $N = 600$. Denote $N_p$ and $N_t$ as the numbers of subjects randomized to placebo and treatment, respectively, and $a_i$ as the group indicator for patient $i$. For illustration, the data collection time for patient $i$ is assumed as $t_i = i$. The standard deviations of two groups are set as $\sigma_p = \sigma_t = 0.3$ as homoscedasticity, and $\sigma_p = 0.4$, $\sigma_t = 0.2$ as heteroscedasticity. The proportion of subjects randomized to treatment is evaluated at $2$ magnitudes of $0.5$ and $0.75$. We consider 4 different function forms of $f(t)$ listed below, where S1 corresponds to a constant placebo effect, S2 is for a linear effect, and S3 and S4 are for non-linear effects with $\lfloor t \rfloor$ as the greatest number less than or equal to $t$. To obtain stable results in S3 and S4, we use different random seeds of generating $f(t)$ in different simulation iterations. The number of simulation iterations is $100,000$. 

\begin{itemize}
	\item S1: $f(t) = 0$
	\item S2: $f(t) = 0.3(t-1)/(N-1)$
	\item S3: $f(t) = \sum_{l=1}^{\lfloor t \rfloor} \eta_l $, where $\eta_l \sim \mathcal{N}(0, {0.002})$
	\item S4: $f(t) = \sum_{l=1}^{\lfloor t \rfloor} \eta_l $, where $\eta_l \sim \mathcal{N}(0, {0.004})$ 
\end{itemize}

All methods for evaluation are listed below. SLR, WLR, RR, RT(T) and RT(W) are briefly reviewed in Section \ref{sec:comparator}. Specifically, RT(T) and RT(W) use the Welch's $t$-test statistic and the Wilcoxon rank sum test statistic, respectively, in their randomization test components. In Step 1 of SWSR, we consider $C=4$ candidate sets of hyperparameters for cross-validation: $k=1, d=1$; $k=1, d=2$; $k=5, d=2$; and $k=5, d=3$. Later results (Table \ref{T:sim_com}) suggest that $k$ should be chosen at a moderate value to avoid deflated type I error rate and power loss. 

\begin{itemize}
	\item WTT: Welch's $t$-test
	\item WT: Wilcoxon rank sum test
	\item SLR: Simple linear regression with intercept, covariates $t_i$ and $a_i$
	\item WLR: Weighted linear regression with intercept, covariates $t_i$ and $a_i$
	\item RR: Robust linear regression with intercept, covariates $t_i$ and $a_i$
	\item RT(T): Randomization test with the Welch's $t$-test statistic
	\item RT(W): Randomization test with the Wilcoxon rank sum test statistic
	\item SWSR: Semiparametric Weighted Spline Regression
\end{itemize}

Note that SLR, WLR, RR specify the correct parametric (linear) form of $t_i$ under S1 and S2. We first evaluate those methods under a homoscedasticity setting of $\sigma_p = \sigma_t = 0.3$ and $N_p = N_t = 300$ with results of type I error rate and power in Table \ref{T:sim_eq_power}, estimation bias and standard error of $\widehat{\theta}$ in Supplementary Material Web Table 1 and coverage probability in Supplementary Material Web Table 2. The first 4 rows in Table \ref{T:sim_eq_power} report type I error rates under 4 scenarios with $\theta = 0$. All methods have relatively accurate type I error rates at the nominal level of $2.5\%$. For power in the next 4 rows with $\theta=0.1$, all methods have a similar performance under S1 and S2. Under S3 and S4, SWSR has the best performance with approximately $10\%$ and $20\%$ power gain versus other methods under S3 and S4, respectively (Table \ref{T:sim_eq_power}). The bias of WTT, WT, SLR, WLR, RR and SWSR is small under all scenarios (Supplementary Material Web Table 1). Overall, SWSR has the smallest standard error, especially under S3 and S4 (Supplementary Material Web Table 1). All evaluated methods have an accurate coverage probability at $95\%$ (Supplementary Material Web Table 2). The point estimation and confidence interval computation of randomization tests RT(T) and RT(W) require more in-depth discussion on definitions, efficient computational algorithms, etc., which are beyond the scope of this work. Interested readers can refer to \citet{wang2020randomization}.

For the setting of heteroscedasticity with $\sigma_p = 0.4$, $\sigma_t = 0.2$, $N_p = 150$, $N_t = 450$, only WTT, WLR, RT(T), and SWSR can properly control type I error rates at $2.5\%$, and therefore their power is further evaluated (Table \ref{T:sim_un_power}). SWSR has a similar power with the other three methods under S1 and S2, but is much more powerful with a power gain of over $10\%$ under S3 and S4 (Table \ref{T:sim_un_power}). The bias of SWSR is small under all scenarios, while other methods have a slightly larger bias under S4 (Supplementary Material Web Table 3). The standard error of SWSR is similar to other methods under S1 and S2, but is smaller than other methods under S3 and S4 (Supplementary Material Web Table 3). Only SWSR, WTT and WLR have accurate coverage probabilities with results in Supplementary Material Web Table 4. 

The general conclusion is that SWSR can properly control type I error rates at the nominal level across all scenarios evaluated, and has an overall best power performance. 

We further conduct another comparison between SWSR in Algorithm \ref{alg_SSR} with cross-validation, SWSR(S) with a simple spline model of $k=1$ and $d=1$, and SWSR(C) with a complex spline model of $k=100$ and $d=3$. Table \ref{T:sim_com} shows that SWSR(S) can be less powerful than SWSR in some settings, e.g., $10\%$ power loss under S4 with equal-variance. SWSR(C) has an inflated type I error rate under unequal-variance, and therefore, its power is not reported for comparison. It is slightly more powerful than SWSR under some settings, e.g., by approximately $2\%$ under equal-variance S4. Overall, SWSR is more stable with a proper type I error rate control, sufficient power, small estimation bias and accurate coverage probability. 

\begin{table}[ht]
	\centering
	\begin{tabular}{llcccccccc}
		\hline
		$\theta$	& Scenario & WTT & WT & SLR & WLR & RR & RT(T) & RT(W) & SWSR \\ 
		\hline
0 & S1 & 2.47 & 2.46 & 2.51 & 2.51 & 2.49 & 2.56 & 2.56 & 2.51 \\ 
& S2 & 2.47 & 2.55 & 2.50 & 2.50 & 2.56 & 2.58 & 2.65 & 2.53 \\ 
& S3 & 2.53 & 2.52 & 2.51 & 2.51 & 2.55 & 2.65 & 2.61 & 2.62 \\ 
& S4 & 2.54 & 2.53 & 2.58 & 2.58 & 2.59 & 2.63 & 2.65 & 2.53 \\ 
\\
0.1 & S1 & 98.27 & 97.84 & 98.27 & 98.27 & 97.77 & 98.27 & 97.86 & 98.22 \\ 
& S2 & 97.39 & 96.81 & 98.23 & 98.23 & 97.72 & 97.42 & 96.88 & 98.17 \\ 
& S3 & 68.07 & 66.57 & 84.45 & 84.47 & 82.31 & 68.40 & 66.81 & 96.15 \\ 
& S4 & 51.26 & 50.78 & 71.03 & 71.03 & 68.53 & 51.63 & 51.10 & 93.62 \\ 
		\hline
	\end{tabular}
	\caption{Type I error rate and power under $\sigma_p = \sigma_t = 0.3$ and $N_p = N_t = 300$. }
	\label{T:sim_eq_power}
\end{table}


\begin{table}[ht]
	\centering
	\begin{tabular}{llcccccccc}
		\hline
		$\theta$	& Scenario & WTT & WT & SLR & WLR & RR & RT(T) & RT(W) & SWSR \\ 
		\hline
0 & S1 & 2.46 & 5.10 & 7.60 & 2.56 & 10.84 & 2.71 & 5.27 & 2.62 \\ 
& S2 & 2.51 & 4.78 & 7.61 & 2.61 & 10.72 & 2.70 & 4.93 & 2.64 \\ 
& S3 & 2.61 & 3.57 & 5.15 & 2.61 & 5.98 & 2.76 & 3.65 & 2.78 \\ 
& S4 & 2.46 & 3.04 & 4.27 & 2.55 & 4.77 & 2.59 & 3.13 & 2.62 \\ 
\\
0.13 & S1 & 96.72 & - & - & 96.83 & - & 96.89 & - & 96.85 \\ 
& S2 & 95.87 & - & - & 96.84 & - & 96.07 & - & 96.87 \\ 
& S3 & 70.81 & - & - & 85.30 & - & 71.23 & - & 95.06 \\ 
& S4 & 55.07 & - & - & 74.08 & - & 55.61 & - & 92.76 \\ 
		\hline
	\end{tabular}
	\caption{Type I error and power under $\sigma_p = 0.4$, $\sigma_t = 0.2$, $N_p = 150$, $N_t = 450$. }
	\label{T:sim_un_power}
\end{table}

\begin{table}[ht]
	\centering
	\scriptsize
	\begin{tabular}{lllccccccccc}
		\hline
		\multicolumn{3}{c}{} & \multicolumn{3}{c}{Type I error / Power} & \multicolumn{3}{c}{Bias $\times 1000$ (SE $\times 100$)} & \multicolumn{3}{c}{CP}  \\ 
		$\sigma$ & $\theta$ & Scenario & SWSR & SWSR(S) & SWSR(C) & SWSR & SWSR(S) & SWSR(C) & SWSR & SWSR(S) & SWSR(C) \\ 
  \hline
  E & 0 & S1 & 2.51 & 2.48 & 2.53 & 0.1 (2.5) & 0.1 (2.4) & 0.1 (2.7) & 95.0 & 95.0 & 95.0 \\ 
  &  & S2 & 2.53 & 2.51 & 2.50 & -0.0 (2.5) & -0.0 (2.4) & 0.0 (2.7) & 95.0 & 95.0 & 95.0 \\ 
  &  & S3 & 2.62 & 2.56 & 2.53 & -0.0 (2.7) & -0.0 (3.1) & -0.0 (2.7) & 94.9 & 95.0 & 94.9 \\ 
  &  & S4 & 2.53 & 2.57 & 2.51 & -0.1 (2.9) & -0.1 (3.6) & 0.0 (2.7) & 94.9 & 94.9 & 95.0 \\ 
  \\
  & 0.1 & S1 & 98.22 & 98.26 & 96.00 & 0.0 (2.5) & 0.0 (2.5) & 0.1 (2.7) & 94.9 & 94.9 & 94.9 \\ 
  &  & S2 & 98.17 & 98.22 & 96.08 & 0.0 (2.5) & -0.0 (2.5) & 0.0 (2.7) & 94.9 & 94.9 & 94.8 \\ 
  &  & S3 & 96.15 & 90.27 & 95.74 & 0.2 (2.7) & 0.2 (3.1) & 0.1 (2.7) & 95.0 & 94.9 & 94.9 \\ 
  &  & S4 & 93.62 & 80.52 & 95.50 & 0.0 (2.9) & -0.1 (3.6) & 0.0 (2.7) & 95.0 & 95.0 & 95.0 \\ 
  \\
  U & 0 & S1 & 2.62 & 2.58 & 3.36 & 0.1 (3.4) & 0.1 (3.4) & 0.0 (3.6) & 94.7 & 94.8 & 93.2 \\ 
  &  & S2 & 2.64 & 2.61 & 3.41 & 0.2 (3.4) & 0.2 (3.4) & 0.3 (3.5) & 94.8 & 94.9 & 93.4 \\ 
  &  & S3 & 2.78 & 2.70 & 3.45 & 0.2 (3.6) & 0.2 (4.0) & 0.3 (3.6) & 94.6 & 94.7 & 93.3 \\ 
  &  & S4 & 2.62 & 2.57 & 3.29 & 0.0 (3.8) & -0.0 (4.5) & 0.1 (3.6) & 94.8 & 94.9 & 93.4 \\ 
  \\
  & 0.13 & S1 & 96.85 & 96.84 & - & 0.1 (3.4) & 0.1 (3.4) & 0.1 (3.5) & 94.7 & 94.8 & 93.3 \\ 
  &  & S2 & 96.87 & 96.85 & - & -0.1 (3.4) & -0.1 (3.4) & 0.0 (3.5) & 94.8 & 94.8 & 93.3 \\ 
  &  & S3 & 95.06 & 89.98 & - & 0.2 (3.6) & 0.1 (4.0) & 0.1 (3.6) & 94.8 & 94.9 & 93.3 \\ 
  &  & S4 & 92.76 & 82.08 & - & -0.1 (3.8) & -0.3 (4.5) & 0.0 (3.6) & 94.9 & 94.9 & 93.5 \\ 
  \hline
	\end{tabular}
	\caption{Comparisons between SWSR in Algorithm \ref{alg_SSR}, SWSR(S) with a simple spline model and SWSR(C) with a complex spline model under settings of equal-variance ``E'' with $\sigma_p = \sigma_t = 0.3$ and $N_p = N_t = 300$, and unequal-variance ``U'' with $\sigma_p = 0.4$, $\sigma_t = 0.2$, $N_p = 150$, $N_t = 450$. }
	\label{T:sim_com}
\end{table}

\section{A case study}
\label{sec:real}

In this section, we consider a Phase 3 clinical trial of evaluating a new treatment versus placebo in treating Hidradenitis Suppurativa (HS), which is a painful, chronic inflammatory skin disease with few options for effective treatment \citep{kimball2016two}. Fluctuations or variabilities are common in HS-related endpoints, mainly due to the natural disease characteristics, variations of evaluation from different physicians, etc \citep{frew2020quantifying, garg2024addressing}. Therefore, we implement our proposed method SWSR and some other feasible methods to adjust potential time-varying placebo effects.

\begin{figure}
	\centering
	\includegraphics[width=0.8\linewidth]{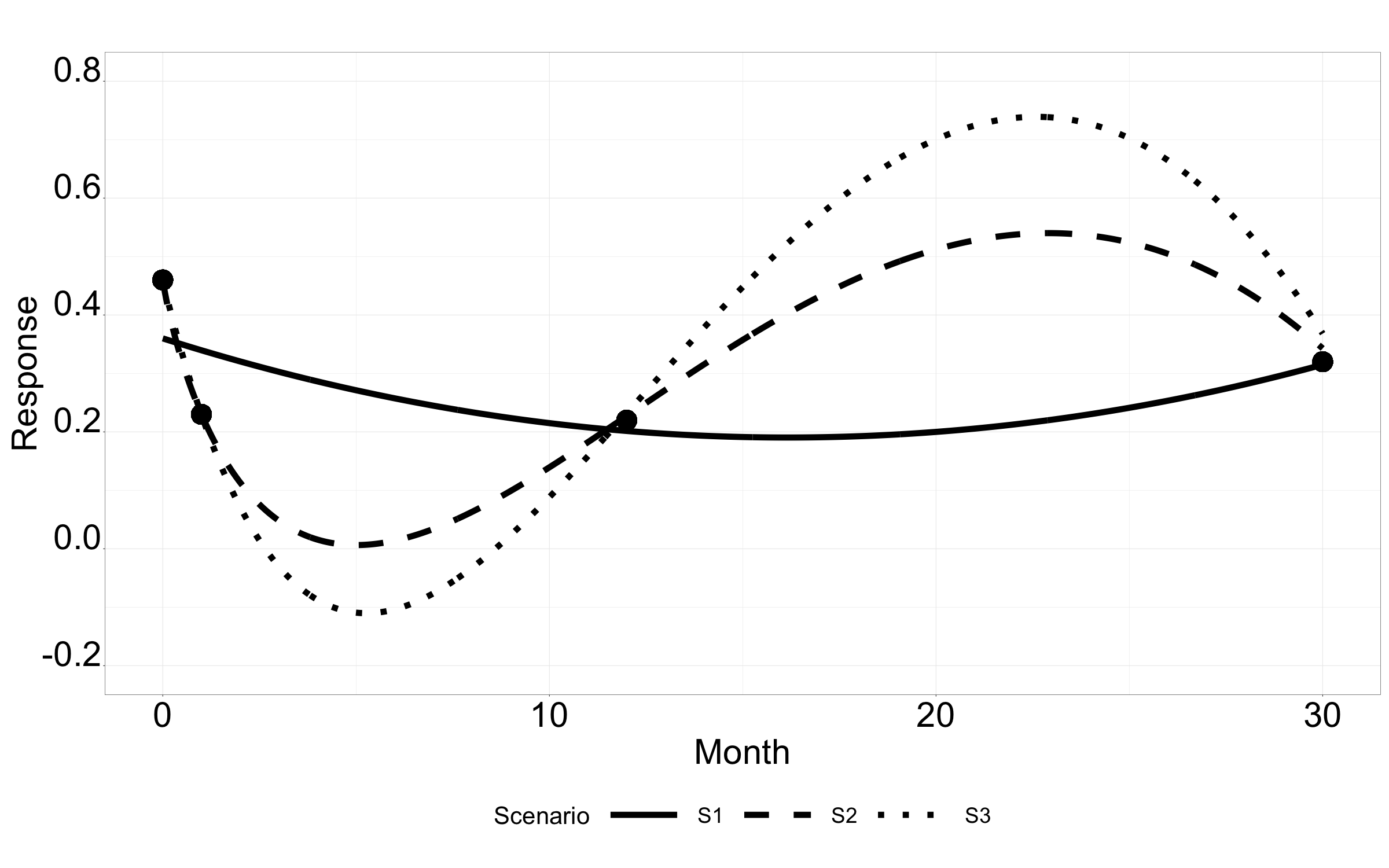}
	\caption{Three different $f(t)$'s as the time-varying placebo effects. }
	\label{fig_real}
\end{figure}

The endpoint of interest is the negative percent change from baseline in total abscess and inflammatory nodule count (PCHG-AN) at Week 16, where a higher value corresponds to a better clinical response. In Table \ref{T:real_assumpt}, we summarize the placebo data of clinical trials from 4 compounds: risankizumab \citep{kimball2023efficacy}, secukinumab \citep{kimball2023secukinumab}, povorcitinib \citep{kirby2023efficacy} and sonelokimab \citep{moonlake}. The relative month $t$ of those trials ranges from $0$ to $30$. Given 4 sets of placebo data, we consider 3 different functional forms of $f(t)$ as the time-varying placebo effects. The coefficients in S1, S2 and S3 are based on least squares estimators of corresponding covariates, e.g., $t$ and $t^2$ for S1. Figure \ref{fig_real} visualizes three curves: S1 represents a flat curve, while S2 and S3 contain more fluctuations. 

\begin{itemize}
	\item S1: $f(t) = 0.36-0.021 \times t+0.00065 \times t^2$
	\item S2: $f(t) = 0.46-0.507 \times t +0.287 \times t^{1.3} -0.00977 \times t^2$
	\item S3: $f(t) = 26.57+0.863 \times t-11.34 \times log(t+10)-0.0114 \times t^2$
\end{itemize}

\begin{table}[ht]
	\centering
	\small
	\begin{tabular}{cccc}
		\hline
		Trials & Calendar Month\textsuperscript{1} & Relative Month\textsuperscript{2} & Mean\textsuperscript{3}\\
		\hline
		Risankizumab Ph2 Trial NCT03926169 & April 2020 & 0 & 0.46\textsuperscript{4} \\
		\\
		Secukinumab Ph3 Trials & May 2020 & 1 & 0.23\textsuperscript{5} \\
		NCT03713619 and NCT03713632 &&&\\
		\\
		Povorcitinib Ph2 Trial NCT04476043 & April 2021 & 12 & 0.22\textsuperscript{4} \\
		\\
		Sonelokimab Ph2 Trial NCT05322473 & October 2022 & 30 & 0.32 \\
		\hline
		\multicolumn{4}{l}{\small{1: The calendar month is based on the middle point of the study start date and the primary completion}} \\
		\multicolumn{4}{l}{\small{date on clinicaltrials.gov.}} \\
		\multicolumn{4}{l}{\small{2: The relative month is with respect to the first Risankizumab Ph2 Trial.}} \\
		\multicolumn{4}{l}{\small{3: The mean of negative PCHG-AN at Week 16, except Week 12 for the Sonelokimab Ph2 Trial.}} \\
		\multicolumn{4}{l}{\small{4: Approximated based on results of CHG-AN and Baseline AN count.}} \\
		\multicolumn{4}{l}{\small{5: A weighted average from 2 trials.}} 
	\end{tabular}
	\caption{Summary of HS trials in 4 compounds. }
	\label{T:real_assumpt}
\end{table}

Suppose that there is another Phase 3 clinical trial conducted during this time interval from $0$ to $30$ with $N_p = N_t = 200$. Both a homoscedasticity of $\sigma_p = \sigma_t = 0.3$ and a heteroscedasticity of $\sigma_p = 0.4$, $\sigma_t = 0.2$ are considered. We evaluate WTT, WLR, RT(T) and SWSR in this study, because they can properly control type I error rates under both homoscedasticity and heteroscedasticity settings based on results in Section \ref{sec:sim}. This feature is desired for practical clinical trials with unknown variance. The number of simulation iterations is $100,000$. 

Table \ref{T:real} shows that all four methods have accurate type I error rates under all settings. SWSR is more powerful than the other three methods under S2 and S3. All methods have a similar power under S1. Across all settings, SWSR has the best overall power, and is especially useful when there are moderate time-varying placebo effects. All methods have small biases of $\widehat{\theta}$ with accurate coverage probabilities. 

\begin{table}[ht]
	\centering
		\footnotesize
	\begin{tabular}{llccccccccccc}
		\hline
		\multicolumn{3}{c}{} & \multicolumn{4}{c}{Type I error / Power} & \multicolumn{3}{c}{Bias $\times 1000$ (SE $\times 100$)} & \multicolumn{3}{c}{CP}  \\ 
		$\theta$ & $\sigma$ & Scenario & WTT & WLR & RT(T) & SWSR & WTT & WLR & SWSR & WTT & WLR & SWSR \\ 
		\hline
0 & E & S1 & 2.44 & 2.49 & 2.54 & 2.47 & -0.0 (3.0) & -0.0 (3.0) & -0.1 (3.0) & 95.0 & 95.0 & 94.9 \\ 
&  & S2 & 2.51 & 2.53 & 2.63 & 2.63 & -0.1 (3.5) & -0.1 (3.2) & -0.1 (3.0) & 94.9 & 95.0 & 94.8 \\ 
&  & S3 & 2.58 & 2.63 & 2.70 & 2.61 & -0.1 (4.3) & 0.0 (3.5) & 0.0 (3.0) & 94.8 & 94.8 & 94.9 \\ 
\\
& U & S1 & 2.50 & 2.54 & 2.60 & 2.60 & -0.0 (3.2) & -0.1 (3.2) & -0.0 (3.2) & 94.8 & 94.8 & 94.7 \\ 
&  & S2 & 2.53 & 2.57 & 2.61 & 2.57 & -0.1 (3.7) & -0.1 (3.3) & -0.2 (3.2) & 94.9 & 94.8 & 94.8 \\ 
&  & S3 & 2.53 & 2.59 & 2.62 & 2.58 & 0.2 (4.3) & 0.1 (3.6) & 0.1 (3.2) & 95.0 & 95.0 & 94.9 \\ 
\\
0.12 & E & S1 & 97.58 & 97.62 & 97.61 & 97.77 & 0.1 (3.0) & 0.1 (3.0) & 0.1 (3.0) & 94.9 & 94.9 & 94.9 \\ 
&  & S2 & 92.06 & 96.30 & 92.08 & 97.67 & -0.2 (3.5) & -0.1 (3.2) & -0.0 (3.0) & 94.9 & 94.8 & 94.9 \\ 
&  & S3 & 80.56 & 93.14 & 80.66 & 97.71 & 0.1 (4.2) & -0.0 (3.5) & 0.0 (3.0) & 94.9 & 95.0 & 95.0 \\
\\ 
& U & S1 & 96.21 & 96.24 & 96.25 & 96.52 & -0.0 (3.2) & -0.1 (3.2) & -0.0 (3.2) & 94.9 & 94.9 & 94.8 \\ 
&  & S2 & 90.03 & 94.73 & 90.20 & 96.28 & -0.1 (3.7) & -0.1 (3.3) & -0.3 (3.2) & 94.9 & 94.9 & 94.7 \\ 
&  & S3 & 78.58 & 91.31 & 78.69 & 96.41 & 0.1 (4.3) & 0.1 (3.6) & -0.0 (3.2) & 95.0 & 94.9 & 94.7 \\ 
		\hline
	\end{tabular}
	\caption{Case study results of WTT, WLR, RT(T) and SWSR under settings of equal-variance ``E'' with $\sigma_p = \sigma_t = 0.3$, and unequal-variance ``U'' with $\sigma_p = 0.4$, $\sigma_t = 0.2$. }
	\label{T:real}
\end{table}

\section{Discussions}
\label{sec:dis}

In this article, we propose SWSR (Semiparametric Weighted Spline Regression) to parametrically estimate the treatment effect $\theta$ with a nonparametric form on the time-varying placebo effects. Based on simulation studies and a case study, our method can properly control the type I error rate with an overall high power across varying settings, and is robust to unknown time-varying placebo effects. 

The time-varying placebo effect in SWSR is modeled by B-splines, and can be generally handled by other nonparametric methods. Supplementary Material Section 2.1 considers an option of P-splines with penalties \citep{eilers1996flexible, pya2015shape}. Simulation results show that B-splines and P-splines in SWSR perform similarly. Additionally, a simulation study with uneven data collection time is conducted in Supplementary Material Section 2.2. The conclusion of different methods is similar to Section \ref{sec:sim}.  

On the type I error rate control of SWSR, Supplementary Material Section 3.1 provides some theoretical discussion of the asymptotic property of SWSR estimator. From a practical perspective, simulation studies are also needed to evaluate finite-sample performance. Supplementary Material Section 3.2 conducts simulations under additional scenarios under the null hypothesis to show that SWSR has a well-controlled type I error rate. 

As specified in (\ref{equ:assump}) and (\ref{equ:mean}), there is no assumed treatment by time interaction effect. The temporal variable $f(t)$ is a so-called prognostic covariate that is associated with the outcome $y$ but not the treatment group indicator $a$ \citep{fda2023}. Exploratory simulation results show that the current SWSR with no treatment by time interaction terms is not robust to potential model misspecification with interaction. Some extensions can be conducted to accommodate the interaction, for example, adding interaction terms between $t$ and $a$ to (\ref{equ:mean}), modeling the time-varying treatment effect $\theta$ by several piecewise components. 

For the internal knots of B-splines in Algorithm \ref{alg_SSR}, we place them evenly on the time axis, e.g., the internal knot is at the middle point when $k=1$. In practical trials, external information can be leveraged to add some customized knots, e.g., before and after the COVID-19 pandemic. Specific strategies can be adapted during trial conduct, but should be finalized before the unblinding process and further analyses. 

As a limitation, our method is computationally intensive with cross-validation to select proper hyperparameters. It can be slightly less powerful than some methods under some specific functional forms of the time-varying placebo effects $f(t)$, but is able to preserve satisfactory power under varying scenarios. Some future works include covariate adjustments in $f(t)$ or additional parametric components related to the treatment effect, generalization to other endpoints (e.g., categorical) and other parametric forms of the treatment effect (e.g., odds ratio).  

\section*{Supplementary Material}

\begin{description}
	\item[Supplementary Document:] This Supplementary Material file contains additional simulation results that are referred to in this article and some theoretical discussion. 
	\item[R code:] The R code is available on the JCGS website and also on Github \url{https://github.com/tian-yu-zhan/SWSR}.
\end{description}

\section*{Acknowledgements}
The authors thank the Editor, the Associate Editor and the reviewers for their constructive comments on improving this article significantly.

This manuscript was supported by AbbVie Inc. AbbVie participated in the review and approval of the content. Tianyu Zhan and Yihua Gu are employed by AbbVie Inc., and may own AbbVie stock. 

\section*{Competing interests}
No competing interest is declared.

%

\bigskip

\bibliographystyle{Chicago}

\bibliography{TD_ref}

\begin{thebibliography}{}

\bibitem[\protect\citeauthoryear{Coad}{Coad}{1991}]{coad1991sequential}
Coad, D. (1991).
\newblock Sequential tests for an unstable response variable.
\newblock {\em Biometrika\/}~{\em 78\/}(1), 113--121.

\bibitem[\protect\citeauthoryear{De~Boor}{De~Boor}{1978}]{de1978practical}
De~Boor, C. (1978).
\newblock {\em A practical guide to splines}.
\newblock Springer-Verlag, New York.

\bibitem[\protect\citeauthoryear{Eilers and Marx}{Eilers and
  Marx}{1996}]{eilers1996flexible}
Eilers, P.~H. and B.~D. Marx (1996).
\newblock {Flexible smoothing with B-splines and penalties}.
\newblock {\em Statistical Science\/}~{\em 11\/}(2), 89--121.

\bibitem[\protect\citeauthoryear{Fahrmeir, Kneib, Lang, and Marx}{Fahrmeir
  et~al.}{2013}]{fahrmeir2013regression}
Fahrmeir, L., T.~Kneib, S.~Lang, and B.~D. Marx (2013).
\newblock {\em Regression: Models, Methods and Applications}.
\newblock Springer.

\bibitem[\protect\citeauthoryear{{Food and Drug Administration}}{{Food and Drug
  Administration}}{2023}]{fda2023}
{Food and Drug Administration} (2023).
\newblock {Adjusting for Covariates in Randomized Clinical Trials for Drugs and
  Biological Products}.
\newblock
  \url{https://www.fda.gov/regulatory-information/search-fda-guidance-documents/adjusting-covariates-randomized-clinical-trials-drugs-and-biological-products}.

\bibitem[\protect\citeauthoryear{Frew, Jiang, Singh, Navrazhina, Vaughan, and
  Krueger}{Frew et~al.}{2020}]{frew2020quantifying}
Frew, J.~W., C.~S. Jiang, N.~Singh, K.~Navrazhina, R.~Vaughan, and J.~G.
  Krueger (2020).
\newblock Quantifying the natural variation in lesion counts over time in
  untreated hidradenitis suppurativa: implications for outcome measures and
  trial design.
\newblock {\em JAAD International\/}~{\em 1\/}(2), 208--221.

\bibitem[\protect\citeauthoryear{Garg, Mastacouris, Ingram, and Strunk}{Garg
  et~al.}{2024}]{garg2024addressing}
Garg, A., N.~Mastacouris, J.~R. Ingram, and A.~Strunk (2024).
\newblock Addressing high placebo response rates in randomized clinical trials
  for hidradenitis suppurativa.
\newblock {\em British Journal of Dermatology\/}~{\em 190\/}(3), 427--429.

\bibitem[\protect\citeauthoryear{Hastie and Tibshirani}{Hastie and
  Tibshirani}{1990}]{hastie1990generalized}
Hastie, T. and R.~Tibshirani (1990).
\newblock {\em Generalized Additive Models}.
\newblock Chapman \& Hall.

\bibitem[\protect\citeauthoryear{Huber and Ronchetti}{Huber and
  Ronchetti}{2011}]{huber2011robust}
Huber, P.~J. and E.~M. Ronchetti (2011).
\newblock {\em Robust statistics}.
\newblock John Wiley \& Sons.

\bibitem[\protect\citeauthoryear{Jennison and Turnbull}{Jennison and
  Turnbull}{2001}]{jennison2001group}
Jennison, C. and B.~W. Turnbull (2001).
\newblock Group sequential tests with outcome-dependent treatment assignment.
\newblock {\em Sequential Analysis\/}~{\em 20\/}(4), 209--234.

\bibitem[\protect\citeauthoryear{Kimball, Jemec, Alavi, Reguiai, Gottlieb,
  Bechara, Paul, Bourboulis, Villani, Schwinn, et~al.}{Kimball
  et~al.}{2023}]{kimball2023secukinumab}
Kimball, A.~B., G.~B. Jemec, A.~Alavi, Z.~Reguiai, A.~B. Gottlieb, F.~G.
  Bechara, C.~Paul, E.~J.~G. Bourboulis, A.~P. Villani, A.~Schwinn, et~al.
  (2023).
\newblock {Secukinumab in moderate-to-severe hidradenitis suppurativa (SUNSHINE
  and SUNRISE): Week 16 and week 52 results of two identical, multicentre,
  randomised, placebo-controlled, double-blind phase 3 trials}.
\newblock {\em The Lancet\/}~{\em 401\/}(10378), 747--761.

\bibitem[\protect\citeauthoryear{Kimball, Okun, Williams, Gottlieb, Papp,
  Zouboulis, Armstrong, Kerdel, Gold, Forman, et~al.}{Kimball
  et~al.}{2016}]{kimball2016two}
Kimball, A.~B., M.~M. Okun, D.~A. Williams, A.~B. Gottlieb, K.~A. Papp, C.~C.
  Zouboulis, A.~W. Armstrong, F.~Kerdel, M.~H. Gold, S.~B. Forman, et~al.
  (2016).
\newblock Two phase 3 trials of adalimumab for hidradenitis suppurativa.
\newblock {\em New England Journal of Medicine\/}~{\em 375\/}(5), 422--434.

\bibitem[\protect\citeauthoryear{Kimball, Prens, Passeron, Maverakis, Turchin,
  Beeck, Drogaris, Geng, Zhan, Messina, et~al.}{Kimball
  et~al.}{2023}]{kimball2023efficacy}
Kimball, A.~B., E.~P. Prens, T.~Passeron, E.~Maverakis, I.~Turchin, S.~Beeck,
  L.~Drogaris, Z.~Geng, T.~Zhan, I.~Messina, et~al. (2023).
\newblock {Efficacy and Safety of Risankizumab for the Treatment of
  Hidradenitis Suppurativa: A Phase 2, Randomized, Placebo-Controlled Trial}.
\newblock {\em Dermatology and Therapy\/}~{\em 13\/}(5), 1099--1111.

\bibitem[\protect\citeauthoryear{Kirby, Okun, Alavi, Bechara, Zouboulis, Brown,
  Santos, Wang, Bibeau, Kimball, et~al.}{Kirby
  et~al.}{2023}]{kirby2023efficacy}
Kirby, J.~S., M.~M. Okun, A.~Alavi, F.~G. Bechara, C.~C. Zouboulis, K.~Brown,
  L.~L. Santos, A.~Wang, K.~B. Bibeau, A.~B. Kimball, et~al. (2023).
\newblock {Efficacy and safety of the oral Janus kinase 1 inhibitor
  povorcitinib (INCB054707) in patients with hidradenitis suppurativa in a
  phase 2, randomized, double-blind, dose-ranging, placebo-controlled study}.
\newblock {\em Journal of the American Academy of Dermatology\/}.

\bibitem[\protect\citeauthoryear{Korn and Freidlin}{Korn and
  Freidlin}{2011}]{korn2011outcome}
Korn, E.~L. and B.~Freidlin (2011).
\newblock Outcome-adaptive randomization: is it useful?
\newblock {\em Journal of Clinical Oncology\/}~{\em 29\/}(6), 771.

\bibitem[\protect\citeauthoryear{Lehmann and Romano}{Lehmann and
  Romano}{2006}]{lehmann2006testing}
Lehmann, E.~L. and J.~P. Romano (2006).
\newblock {\em Testing Statistical Hypotheses}.
\newblock Springer Science \& Business Media.

\bibitem[\protect\citeauthoryear{Liu and Lee}{Liu and
  Lee}{2015}]{liu2015overview}
Liu, S. and J.~J. Lee (2015).
\newblock {An overview of the design and conduct of the BATTLE trials}.
\newblock {\em Chinese Clinical Oncology\/}~{\em 4\/}(3).

\bibitem[\protect\citeauthoryear{Marsh and Cormier}{Marsh and
  Cormier}{2001}]{marsh2001spline}
Marsh, L.~C. and D.~R. Cormier (2001).
\newblock {\em Spline Regression Models}.
\newblock Sage.

\bibitem[\protect\citeauthoryear{Montgomery, Peck, and Vining}{Montgomery
  et~al.}{2021}]{montgomery2021introduction}
Montgomery, D.~C., E.~A. Peck, and G.~G. Vining (2021).
\newblock {\em Introduction to linear regression analysis}.
\newblock John Wiley \& Sons.

\bibitem[\protect\citeauthoryear{MoonLake}{MoonLake}{2023}]{moonlake}
MoonLake (2023).
\newblock {MoonLake Immunotherapeutics R\&D Day Webcast Presentation Document
  Results MIRA trial}.
\newblock
  \url{https://ir.moonlaketx.com/static-files/ee61d19d-63d5-4595-98e6-316a814da0a4}.

\bibitem[\protect\citeauthoryear{Proschan and Evans}{Proschan and
  Evans}{2020}]{proschan2020resist}
Proschan, M. and S.~Evans (2020).
\newblock Resist the temptation of response-adaptive randomization.
\newblock {\em Clinical Infectious Diseases\/}~{\em 71\/}(11), 3002--3004.

\bibitem[\protect\citeauthoryear{Proschan and Dodd}{Proschan and
  Dodd}{2019}]{proschan2019re}
Proschan, M.~A. and L.~E. Dodd (2019).
\newblock Re-randomization tests in clinical trials.
\newblock {\em Statistics in Medicine\/}~{\em 38\/}(12), 2292--2302.

\bibitem[\protect\citeauthoryear{Pya and Wood}{Pya and
  Wood}{2015}]{pya2015shape}
Pya, N. and S.~N. Wood (2015).
\newblock Shape constrained additive models.
\newblock {\em Statistics and computing\/}~{\em 25}, 543--559.

\bibitem[\protect\citeauthoryear{Robertson, Lee, L{\'o}pez-Kolkovska, and
  Villar}{Robertson et~al.}{2023}]{robertson2023response}
Robertson, D.~S., K.~M. Lee, B.~C. L{\'o}pez-Kolkovska, and S.~S. Villar
  (2023).
\newblock Response-adaptive randomization in clinical trials: from myths to
  practical considerations.
\newblock {\em Statistical Science\/}~{\em 38\/}(2), 185.

\bibitem[\protect\citeauthoryear{Roig, Krotka, Burman, Glimm, Gold, Hees,
  Jacko, Koenig, Magirr, Mesenbrink, et~al.}{Roig et~al.}{2022}]{roig2022model}
Roig, M.~B., P.~Krotka, C.-F. Burman, E.~Glimm, S.~M. Gold, K.~Hees, P.~Jacko,
  F.~Koenig, D.~Magirr, P.~Mesenbrink, et~al. (2022).
\newblock On model-based time trend adjustments in platform trials with
  non-concurrent controls.
\newblock {\em BMC Medical Research Methodology\/}~{\em 22\/}(1), 1--16.

\bibitem[\protect\citeauthoryear{Rosenberger and Lachin}{Rosenberger and
  Lachin}{2015}]{rosenberger2015randomization}
Rosenberger, W.~F. and J.~M. Lachin (2015).
\newblock {\em Randomization in clinical trials: theory and practice}.
\newblock John Wiley \& Sons.

\bibitem[\protect\citeauthoryear{Shao}{Shao}{2008}]{shao2008mathematical}
Shao, J. (2008).
\newblock {\em Mathematical statistics}.
\newblock Springer Science \& Business Media.

\bibitem[\protect\citeauthoryear{Simon and Simon}{Simon and
  Simon}{2011}]{simon2011using}
Simon, R. and N.~R. Simon (2011).
\newblock {Using randomization tests to preserve type I error with response
  adaptive and covariate adaptive randomization}.
\newblock {\em Statistics \& Probability Letters\/}~{\em 81\/}(7), 767--772.

\bibitem[\protect\citeauthoryear{Venables and Ripley}{Venables and
  Ripley}{2002}]{ven2002}
Venables, W.~N. and B.~D. Ripley (2002).
\newblock {\em Modern Applied Statistics with S\/} (Fourth ed.).
\newblock New York: Springer.
\newblock ISBN 0-387-95457-0.

\bibitem[\protect\citeauthoryear{Villar, Bowden, and Wason}{Villar
  et~al.}{2018}]{villar2018response}
Villar, S.~S., J.~Bowden, and J.~Wason (2018).
\newblock Response-adaptive designs for binary responses: how to offer patient
  benefit while being robust to time trends?
\newblock {\em Pharmaceutical Statistics\/}~{\em 17\/}(2), 182--197.

\bibitem[\protect\citeauthoryear{Wang, Lin, Rosner, and Soon}{Wang
  et~al.}{2023}]{wang2023bayesian}
Wang, C., M.~Lin, G.~L. Rosner, and G.~Soon (2023).
\newblock {A Bayesian model with application for adaptive platform trials
  having temporal changes}.
\newblock {\em Biometrics\/}~{\em 79\/}(2), 1446--1458.

\bibitem[\protect\citeauthoryear{Wang and Rosenberger}{Wang and
  Rosenberger}{2020}]{wang2020randomization}
Wang, Y. and W.~F. Rosenberger (2020).
\newblock Randomization-based interval estimation in randomized clinical
  trials.
\newblock {\em Statistics in Medicine\/}~{\em 39\/}(21), 2843--2854.

\bibitem[\protect\citeauthoryear{Wong, Siah, and Lo}{Wong
  et~al.}{2019}]{wong2019estimation}
Wong, C.~H., K.~W. Siah, and A.~W. Lo (2019).
\newblock Estimation of clinical trial success rates and related parameters.
\newblock {\em Biostatistics\/}~{\em 20\/}(2), 273--286.

\end{thebibliography}
\end{document}